\documentclass[dvips]{article}
\usepackage{supertabular,lscape,epsfig}
\usepackage{amssymb}
\usepackage{amsmath}
\usepackage{rotating}
\usepackage{booktabs}
\DeclareSymbolFont{ppa}{OT1}{ppl}{m}{it}
\DeclareMathSymbol{\vv}{\mathalpha}{ppa}{'166}

\begin{document}

\newcommand{\dd}{\,{\rm d}}
\newcommand{\ie}{{\it i.e.},\,}
\newcommand{\etal}{{\it et al.\ }}
\newcommand{\eg}{{\it e.g.},\,}
\newcommand{\cf}{{\it cf.\ }}
\newcommand{\vs}{{\it vs.\ }}
\newcommand{\zdot}{\makebox[0pt][l]{.}}
\newcommand{\up}[1]{\ifmmode^{\rm #1}\else$^{\rm #1}$\fi}
\newcommand{\dn}[1]{\ifmmode_{\rm #1}\else$_{\rm #1}$\fi}
\newcommand{\upd}{\up{d}}
\newcommand{\uph}{\up{h}}
\newcommand{\upm}{\up{m}}
\newcommand{\ups}{\up{s}}
\newcommand{\arcd}{\ifmmode^{\circ}\else$^{\circ}$\fi}
\newcommand{\arcm}{\ifmmode{'}\else$'$\fi}
\newcommand{\arcs}{\ifmmode{''}\else$''$\fi}
\newcommand{\MS}{{\rm M}\ifmmode_{\odot}\else$_{\odot}$\fi}
\newcommand{\RS}{{\rm R}\ifmmode_{\odot}\else$_{\odot}$\fi}
\newcommand{\LS}{{\rm L}\ifmmode_{\odot}\else$_{\odot}$\fi}
\newcommand{\feh}{\hbox{$ [{\rm Fe}/{\rm H}]$}}

\newcommand{\Abstract}[2]{{\footnotesize\begin{center}ABSTRACT\end{center}
\vspace{1mm}\par#1\par
\noindent
{~}{\it #2}}}

\newcommand{\TabCap}[2]{\begin{center}\parbox[t]{#1}{\begin{center}
  \small {\spaceskip 2pt plus 1pt minus 1pt T a b l e}
  \refstepcounter{table}\thetable \\[2mm]
  \footnotesize #2 \end{center}}\end{center}}

\newcommand{\TableSep}[2]{\begin{table}[p]\vspace{#1}
\TabCap{#2}\end{table}}

\newcommand{\FigCap}[1]{\footnotesize\par\noindent Fig.\  %                     
  \refstepcounter{figure}\thefigure. #1\par}

\newcommand{\TableFont}{\footnotesize}
\newcommand{\TableFontIt}{\ttit}
\newcommand{\SetTableFont}[1]{\renewcommand{\TableFont}{#1}}

\newcommand{\MakeTable}[4]{\begin{table}[htb]\TabCap{#2}{#3}
  \begin{center} \TableFont \begin{tabular}{#1} #4
  \end{tabular}\end{center}\end{table}}

\newcommand{\MakeTableSep}[4]{\begin{table}[p]\TabCap{#2}{#3}
  \begin{center} \TableFont \begin{tabular}{#1} #4
  \end{tabular}\end{center}\end{table}}

\newenvironment{references}%                                                    
{
\footnotesize \frenchspacing
\renewcommand{\thesection}{}
\renewcommand{\in}{{\rm in }}
\renewcommand{\AA}{Astron.\ Astrophys.}
\newcommand{\AAS}{Astron.~Astrophys.~Suppl.~Ser.}
\newcommand{\ApJ}{Astrophys.\ J.}
\newcommand{\ApJS}{Astrophys.\ J.~Suppl.~Ser.}
\newcommand{\ApJL}{Astrophys.\ J.~Letters}
\newcommand{\AJ}{Astron.\ J.}
\newcommand{\IBVS}{IBVS}
\newcommand{\PASJ}{PASJ}
\newcommand{\PASP}{P.A.S.P.}
\newcommand{\Acta}{Acta Astron.}
\newcommand{\MNRAS}{MNRAS}
\renewcommand{\and}{{\rm and }}
\section{{\rm REFERENCES}}
\sloppy \hyphenpenalty10000
\begin{list}{}{\leftmargin1cm\listparindent-1cm
\itemindent\listparindent\parsep0pt\itemsep0pt}}%                               
{\end{list}\vspace{2mm}}

\def\TYLDA{~}
\newlength{\DW}
\settowidth{\DW}{0}
\newcommand{\dw}{\hspace{\DW}}

\newcommand{\refitem}[5]{\item[]{#1} #2%                                        
\def\REFARG{#3}\ifx\REFARG\TYLDA\else, {\it#3}\fi
\def\REFARG{#4}\ifx\REFARG\TYLDA\else, {\bf#4}\fi
\def\REFARG{#5}\ifx\REFARG\TYLDA\else, {#5}\fi.}

\newcommand{\Section}[1]{\section{#1}}
\newcommand{\Subsection}[1]{\subsection{#1}}
\newcommand{\Acknow}[1]{\par\vspace{5mm}{\bf Acknowledgments.} #1}
\pagestyle{myheadings}

\newfont{\bb}{ptmbi8t at 12pt}
\newcommand{\xrule}{\rule{0pt}{2.5ex}}
\newcommand{\xxrule}{\rule[-1.8ex]{0pt}{4.5ex}}
\def\thefootnote{\fnsymbol{footnote}}

\begin{center}
{\Large\bf Blue Large-Amplitude Pulsators and Other Short-Period Variable Stars in the OGLE-IV Galactic Disk Fields}
\vskip1cm
{\bf
J.~~B~o~r~o~w~i~c~z$^1$,~~P.~~P~i~e~t~r~u~k~o~w~i~c~z$^1$,~~P.~~M~r~\'o~z$^1$,~~I.~~S~o~s~z~y~\'n~s~k~i$^1$,\\
A.~~U~d~a~l~s~k~i$^1$,~~M.~K.~~S~z~y~m~a~\'n~s~k~i$^1$,~~K.~~U~l~a~c~z~y~k$^2$,\\
R.~~P~o~l~e~s~k~i$^1$,~~S.~~K~o~z~{\l}~o~w~s~k~i$^1$,~~J.~~S~k~o~w~r~o~n$^1$,\\
D.~M.~~S~k~o~w~r~o~n$^1$,~~K.~~R~y~b~i~c~k~i$^{1,3}$,~~P.~~I~w~a~n~e~k$^1$,\\
M.~~W~r~o~n~a$^1$,~~and~~M.~~G~r~o~m~a~d~z~k~i$^1$\\}
\vskip3mm
{${}^1$ Astronomical Observatory, University of Warsaw, Al. Ujazdowskie 4, 00-478 Warszawa, Poland \\
${}^2$ Department of Physics, University of Warwick, Coventry CV4 7AL, UK \\
${}^3$ Department of Particle Physics and Astrophysics, Weizmann Institute of Science, Rehovot 76100, Israel \\}
\end{center}

\Abstract{Blue Large-Amplitude Pulsators (BLAPs) form a mysterious class of variable stars with typical periods of tens of minutes and amplitudes above 0.1~mag. In this work, we present results of a variability search focused on timescales shorter than 1~h, conducted in OGLE-IV Galactic disk fields containing about 1.1 billion stellar sources down to $I\approx20$~mag. Twenty-five BLAPs have been detected, 20 of which are new discoveries. Their periods range from 8.4 to 62.1~min. We have also found six new eclipsing binary systems with orbital periods from 38.3 to 121.3~min and five short-period large-amplitude ($>0.17$~mag in the $I$-band) variable stars of unknown type.}

{Stars: oscillations (including pulsations) --- Stars: variables: Blue Large-Amplitude Pulsators --- binaries: eclipsing}

\section{Introduction}

Short-period variable stars are intriguing objects that exhibit brightness variations on timescales from seconds to several hours. The variables encompass various types of objects including single as well as binary stars, such as $\delta$~Sct and SX Phe-type pulsating stars, ZZ Cet and V777 Her-type oscillating white dwarfs, variable hot subdwarfs of V361~Hya and V1093 Her type, rapidly oscillating Ap (roAp) stars, Blue Large-Amplitude Pulsators (BLAPs), AM CVn-type binary systems, among others. Numerous short-period variables have been found over the last few years, including pulsating hot subdwarfs (\eg Kupfer \etal 2021), pulsating white dwarfs (\eg Guidry \etal 2021), compact binary systems (\eg Burdge \etal 2019, 2020, Ren \etal 2023), and other low-amplitude stellar sources (\eg Toma \etal 2016, Lin \etal 2023b).

In our work, we focus on the detection and classification of short-period variable stars in the southern part of the Milky Way's stripe observed by the Optical Gravitational Lensing Experiment (OGLE). Although, the OGLE disk fields have been surveyed for variable objects such as Cepheids (Udalski \etal 2018), RR Lyr (Soszy{\'n}ski \etal 2019), and Mira-type stars (Iwanek \etal 2022), those studies focused on longer periods (roughly from hours to months). In the collection of Galactic disk $\delta$~Sct-type stars (Soszy{\'n}ski \etal 2021), there are seven entries with periods below 60~min, with the shortest reported value of 50.2~min. Our goal is an extensive search of the previously unexplored variability range ($<1$~h) to find the mysterious BLAPs and compact binary systems.

BLAPs are a recently discovered class of pulsating stars (Pietrukowicz \etal 2017). They were identified thanks to regular, high-precision photometric observations of hundreds of millions of stars by the OGLE survey (Udalski \etal 2015). BLAPs can be characterized by significant amplitudes of brightness variations, typically 0.1--0.4~mag in optical bands, at very short periods, in the range 3--75~min (Pietrukowicz \etal 2023, submitted). The variability in the stars is attributed to fundamental-mode radial pulsations. Light curve shapes are very similar to those of Cepheids and RR Lyr-type stars, but at much higher temperatures. BLAPs have effective temperatures of $T_{\rm{eff}}$ from about 25~000~K to 34~000~K and surface gravities of $\log g$ from about 4.2 to 5.7~dex. They exhibit a reduced hydrogen content in their atmospheres, compared to main-sequence stars. Such a set of atypical properties makes it difficult to assign specific evolutionary stage to these stars.

It has been proposed that BLAPs are low-mass stars in the pre-white dwarf stage, during which they produce energy in residual hydrogen burning over the degenerate helium core (Romero \etal 2018). Other models indicate that the stars could be in the phase of helium burning in their cores (Byrne and Jeffery 2018, Wu and Li 2018). A third possibility is that BLAPs are shell helium-burning stars with the CO core (Lin \etal 2023a). The stability of pulsation periods ($|1/P \times \Delta P/ \Delta t| < 10^{-6}$ yr$^{-1}$) suggests that these objects evolve on the nuclear timescale. Known BLAPs reside in the Milky Way's bulge and disk. They seem to be absent in the Galactic halo, globular clusters, and Magellanic Clouds (Pietrukowicz 2018). Proposed mechanisms describing the evolutionary history of BLAPs include interactive binary evolution (common envelope phase and/or Roche lobe overflow, Byrne \etal 2021), a close flyby of the supermassive black hole in the center of the Galaxy (Pietrukowicz \etal 2017), and a result (as surviving companions) of single-degenerate Type Ia supernovae (Meng \etal 2020). BLAPs are of particular interest because their pulsations can provide insight into the structure and physical processes in hot (Z-bump) pulsators (Byrne and Jeffery 2018).

So far, over 60 BLAPs have been found, mainly by large-scale surveys. After the discovery of the first 14 pulsators by OGLE (Pietrukowicz \etal 2017), four ultra short-period (3--8 min) BLAPs, characterized by higher gravities (hence named HG-BLAPs), were found in the Zwicky Transient Facility (ZTF; Bellm \etal 2019) data by Kupfer \etal (2019). Recently, 22 candidates for BLAPs from ZTF and four candidates from the OmegaWhite survey (Macfarlane \etal 2015) were proposed by McWhirter and Lam (2022) and Ramsay \etal (2022), respectively. At the moment of writing, another 23 objects from the OGLE Galactic bulge fields can be added to the list of known pulsators (Pietrukowicz \etal 2023, submitted). Finally, first two BLAPs residing in binary systems have been lately reported. Using data from the TESS satellite Pigulski \etal (2022) discovered a pulsator orbiting a B-type main-sequence star HD 133729 every 23.08 days. Based on observations collected with the Tsinghua University-Ma Huateng Telescopes for Survey (TMTS; Lin \etal 2022), Lin \etal (2023a) suggested that object TMTS-BLAP-1 is in a 1576-day binary.

This paper is organized as follows. In Sec.~2, we briefly present the OGLE survey and the collected data. In Sec.~3, we describe the data analysis. The detected short-period variables are listed in Sec.~4. We summarize our results in Sec.~5.

\section{Observations}

The photometric data used for the variability search comes from the fourth phase of the OGLE project (OGLE-IV) which started in March 2010. OGLE has been at the forefront of large-scale photometric sky surveys for over 30 years. The project monitors the southern sky using the 1.3-m Warsaw telescope located at the Las Campanas Observatory in Chile\footnote{The observatory is operated by the Carnegie Institution for Science.}. The most frequently observed regions are the Galactic bulge and Magellanic Clouds. Shallow and less frequent observations of the Galactic disk are conducted since 2013. Currently, the total area observed by the project reaches almost 3600 square degrees. Overall, there are 1338 OGLE-IV disk fields with negative Galactic longitude (GD fields), labeled from GD1063 to GD2475 (with some gaps in numbering). Slightly different nomenclature is followed for 354 disk fields with positive Galactic longitudes. These span from DG1000 to DG1364 (no field numbers between DG1288 and DG1300). The OGLE-IV images are taken using a 268.8-megapixel mosaic camera with a pixel size of $0\zdot\arcs26$ and a 1.4-square-degree field of view, consisting of 32 single 2K$\times$4K CCD detectors. Two filters are used, $V$ and $I$, with the majority of observations taken in the $I$-band. The $I$ filter closely resembles the standard Kron-Cousins filter system, whereas the $V$ filter differs from the standard system to a greater extent. Fields in the Galactic disk are monitored with exposure times of 25~s, while for fields in the inner bulge, it is 100~s. Typically for the GD fields, around 100--200 regular $I$-band measurements and several $V$-band measurements have been obtained. For the DG fields, 10--130 $I$-band observations were taken, but no time-series $V$-band photometry is available. The brightness range in $I$ is between 10.5 and 20~mag. In Fig.~1, we show accuracy of the $I$-band observations for a selected GD field. In total, there are approximately 1.1 billion stellar sources over the analyzed area of about 2400 square degrees of the Galactic disk. More details on the OGLE-IV survey can be found in Udalski \etal (2015). 

\begin{figure}[b!]
\includegraphics[width=12.71cm]{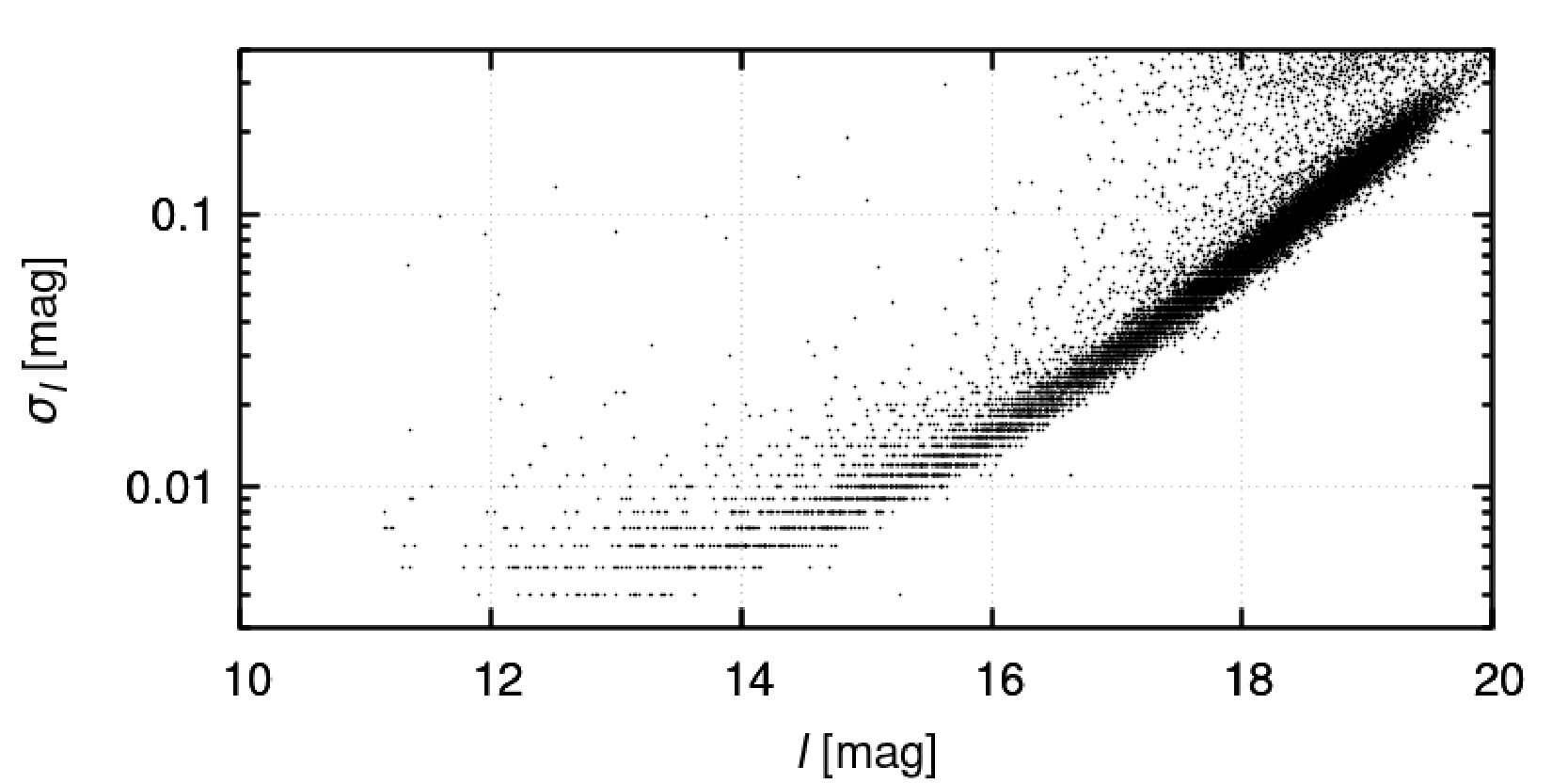}
\FigCap{Accuracy of the I-band photometry in one of the OGLE-IV Galactic disk subfields (GD1915.23).}
\end{figure}

From January to March 2020, additional 3-hour high-cadence observing runs were carried out for selected dense GD fields, resulting in about 100 additional $I$-band measurements per field. This was done for 31 pairs of fields, observed alternately. Shortly after, on 18 March 2020, observations were suspended due to the COVID-19 pandemic. OGLE resumed regular monitoring of the sky on 12 August 2022.

\section{Data Analysis}

Our search for short-period variable stars was conducted with the $I$-band time-series data. For the analysis, we selected sources brighter than $I=20$ mag and with more than 50 measurements. Several DG fields were not included due to small number of observations. The first step was the search for periodicity in all sources using the \textit{FNPEAKS}\footnote[1]{http://helas.astro.uni.wroc.pl/deliverables.php?active=fnpeaks\&lang=en} program, which performs a discrete Fourier transform on unevenly spaced data. We probed a frequency range of $0<f<1000$ cycles per day ($P~>~1.44$~min). Subsequently, only objects with a significant signal-to-noise ratio were selected. The variability signal varied from field to field. The 3-hour high-cadence sequences were searched for periodic signals separately.

In the next step, all objects with periods shorter than 0.05~d were visually inspected. We cleaned our lists of candidate variables from artifacts (objects with fake variability). Obvious outlying points from the light curves of real variables were removed. About 1100 variable stars with periods below 1~h were detected. Stars with amplitudes higher than 0.1 mag were a subject of further inspection. We obtained a set of BLAP candidates based on morphology of the light curves. Variables with low amplitudes were not classified due to difficulties in assigning them to a definitive type. The majority of these objects are likely low-amplitude $\delta$~Sct stars, similar to those found in the Galactic bulge and recently confirmed spectroscopically (Pietrukowicz \etal 2022). Variables exhibiting peculiar light curve shapes after doubling the period were also noted.

We verified the positions of the objects in color-magnitude diagrams (CMDs). The diagrams were constructed for stars from the corresponding OGLE-IV subfields. For objects without $V$-band measurements, we downloaded available data from other sources. In the case of stars from DG fields, we downloaded time-series data in $r$ and $g$-bands from the ZTF survey. For objects from GD fields for which there were no available $V$-band measurements, we used data from the DECam Plane Survey (Schlafly \etal 2018). The final list of BLAPs was created based on the light curve morphology, period, amplitude, and position on the CMD. Moments of the observations were transformed from heliocentric Julian date (HJD) to barycentric Julian date (BJD$_{\rm TDB}$) and the periods were fine-tuned using the \textit{TATRY} code (Schwarzenberg-Czerny 1996). All of the detected objects were checked for additional periodicities.

\section{Results}

The main outcome of this paper are the lists of BLAPs, close eclipsing binary stars, and other short-period variables identified in the OGLE-IV Galactic disk fields. Here, we present phase-folded light curves, along with CMDs, for all the discovered objects. The time-series data are publicly available in the OGLE Collection of Variable Stars (OCVS):
\begin{center}
\textit{https://www.astrouw.edu.pl/ogle/ogle4/OCVS/} 
\end{center}
In the case of BLAPs, we follow the nomenclature used for previously classified variable stars in the OGLE fields. We continue numbering the new objects in fasion OGLE-BLAP-NNN, starting from NNN=038.

\subsection{BLAPs}

The majority of detected short-period variables are of BLAP type. Among the 25 identified BLAPs, 20 objects are new discoveries. Of the five known BLAPs are the prototype and four objects from McWhirter and Lam (2021). We note that nine of the 25 detected BLAPs were classified as variable sources in Gaia DR3 (Eyer \etal 2023), but only three of them belong to the "S" class (short timescale objects). The remaining six stars were classified by Gaia to either the "DSCT|GDOR|SXPHE" class (four stars), or to the "ECL" class (one object), or to the "RS" class (RS~CVn-type, one object). Only for the new variable OGLE-BLAP-039 from the "S" class, Gaia provides the correct period.

Observational parameters, such as coordinates, pulsation periods, magnitudes, and amplitudes of the detected BLAPs are listed in Table~1. Parallaxes $\varpi$ are taken from Gaia EDR3 catalog (Gaia Collaboration 2021). In Fig.~2, we present the period--amplitude diagram with positions of previously discovered BLAPs by OGLE and BLAPs detected in this work. For comparison, the figure also includes single-mode $\delta$~Sct-type stars from the OCVS in the Galactic disk (Soszy{\'n}ski \etal 2021). There are no evident differences in location in the period--amplitude diagram between the objects observed toward the disk and bulge fields. Phase-folded $I$-band light curves and $V-I$ \vs $I$ CMDs for 18 BLAPs detected in the OGLE-IV Galactic disk fields with negative longitudes (GD fields) are shown in Figs.~3--6. For seven pulsators from the OGLE-IV disk fields with positive Galactic longitudes (DG fields), for which no $V$-band measurements are available, the light curves and CMDs based on ZTF data are provided next to the OGLE $I$-band light curves (Figs. 7 and 8). All the OGLE light curves are based on $I$-band measurements, whereas to construct ZTF light curves, we used either $g$ or $r$-band data. In the case of the CMD for object OGLE-BLAP-045, we used the DECam Plane Survey data in the $g$ and $i$ filters.

\begin{sidewaystable}
\caption{Photometric properties of detected BLAPs}
\label{}
{\scriptsize
\begin{tabular}{ccrcrccrccccc}
\hline
Name & RA (J2000) & \multicolumn{1}{c}{Dec (J2000)} & $l$ & \multicolumn{1}{c}{$b$} & $\varpi$ & $\sigma_\varpi$ & \multicolumn{1}{c}{$\textit{P}_{\rm puls}$} & $\sigma_P$ & $I$ & $V-I$ & $A_{I}$ & Other name \\
& [h:m:s] & \multicolumn{1}{c}{[${}^\circ:':''$]} & [${}^\circ$] & \multicolumn{1}{c}{[${}^\circ$]} & [mas] & [mas] & \multicolumn{1}{c}{[min]} & {\tiny [$ 10^{-7}$ min]} & [mag] & [mag] & [mag] & [Ref.] \\ \midrule
OGLE-BLAP-001 & 10:41:48.77 & $-$61:25:08.5 & 288.06356 & $-$2.34712 & 0.14 & 0.07 & 28.2549636 & 33 & 17.217 & 0.335 & 0.194 & {[}1{]} \\ 
OGLE-BLAP-038 & 07:24:10.65 & $-$28:10:11.8 & 241.61485 & $-$5.94461 & --- & --- & 11.2601526 & 60 & 18.822 & 0.844 & 0.413 & --- \\
OGLE-BLAP-039 & 11:00:12.91 & $-$61:13:44.9 & 289.95454 & $-$1.19158 & 0.21 & 0.03 & 31.9651131 & 19 & 15.835 & 0.285 & 0.242 & --- \\
OGLE-BLAP-040 & 11:01:57.46 & $-$65:05:53.2 & 291.73311 & $-$4.63443 & 0.26 & 0.02 & 31.4607770 & 24 & 15.178 & 0.337 & 0.243 & --- \\
OGLE-BLAP-041 & 11:24:36.52 & $-$58:45:58.5 & 291.86639 & 2.22748 & 0.25 & 0.03 & 53.5405245 & 115 & 15.371 & 0.390 & 0.191 & --- \\
OGLE-BLAP-042 & 11:55:36.75 & $-$64:26:53.4 & 296.95618 & $-$2.22606 & 0.33 & 0.09 & 62.0505860 & 96 & 16.993 & 1.046 & 0.147 & --- \\
OGLE-BLAP-043 & 12:04:34.02 & $-$62:12:29.7 & 297.49826 & 0.16800 & 0.20 & 0.22 & 19.0642504 & 14 & 18.125 & 1.505 & 0.342 & --- \\
OGLE-BLAP-044 & 12:29:31.78 & $-$60:48:46.1 & 300.26251 & 1.94534 & 0.54 & 0.03 & 8.4216381 & 6 & 15.713 & 0.189 & 0.211 & --- \\
OGLE-BLAP-045 & 13:04:41.30 & $-$62:21:23.4 & 304.46825 & 0.47572 & 0.23 & 0.07 & 41.3305960 & 109 & 16.457 & --- & 0.191 & --- \\
OGLE-BLAP-046 & 13:08:15.31 & $-$64:39:09.3 & 304.73142 & $-$1.83961 & 0.04 & 0.08 & 39.4729321 & 91 & 17.065 & 0.782 & 0.207 & --- \\
OGLE-BLAP-047 & 15:25:53.96 & $-$52:04:54.8 & 325.53868 & 3.87749 & 0.12 & 0.06 & 31.2450830 & 32 & 16.032 & 0.664 & 0.239 & --- \\
OGLE-BLAP-048 & 15:45:16.20 & $-$59:21:46.0 & 323.49358 & $-$3.61890 & 0.38 & 0.03 & 28.8555579 & 59 & 15.041 & 0.276 & 0.296 & --- \\
OGLE-BLAP-049 & 16:00:21.31 & $-$57:49:26.4 & 325.99337 & $-$3.67615 & 0.31 & 0.08 & 16.4015624 & 7 & 16.597 & 0.367 & 0.375 & --- \\
OGLE-BLAP-050 & 16:21:04.56 & $-$57:33:03.0 & 328.21106 & $-$5.36464 & 0.27 & 0.03 & 22.9352027 & 33 & 15.516 & 0.106 & 0.185 & --- \\
OGLE-BLAP-051 & 16:25:12.57 & $-$47:39:58.4 & 335.68801 & 1.14111 & 0.37 & 0.06 & 45.4160893 & 40 & 15.704 & 1.046 & 0.257 & --- \\
OGLE-BLAP-052 & 16:31:34.60 & $-$53:31:30.8 & 332.15103 & $-$3.63874 & 0.22 & 0.03 & 34.9622342 & 79 & 15.284 & 0.267 & 0.194 & --- \\
OGLE-BLAP-053 & 16:31:35.55 & $-$51:35:22.9 & 333.56826 & $-$2.31842 & --- & --- & 25.2601420 & 76 & 18.323 & 1.180 & 0.330 & --- \\
OGLE-BLAP-054 & 17:17:15.31 & $-$49:09:47.8 & 340.03800 & $-$6.45954 & --- & --- & 11.3324455 & 71 & 17.402 & 0.227 & 0.335 & --- \\
OGLE-BLAP-055 & 18:37:34.76 & $-$04:22:27.0 & 27.45919 & 1.08009 & 0.17 & 0.26 & 35.3945271 & 161 & 17.691 & --- & 0.330 & --- \\
OGLE-BLAP-056 & 18:46:49.17 & $-$10:44:27.7 & 22.83623 & $-$3.86251 & 0.29 & 0.16 & 11.6725445 & 17 & 17.845 & --- & 0.176 & --- \\
OGLE-BLAP-057 & 18:47:45.81 & $-$10:26:06.0 & 23.21496 & $-$3.93102 & 0.13 & 0.05 & 54.6530265 & 114 & 15.752 & --- & 0.112 & ZGP-BLAP-05$^{[2]}$ \\
OGLE-BLAP-058 & 18:52:45.11 & $-$05:10:47.2 & 28.46814 & $-$2.65077 & 0.22 & 0.11 & 35.8390155 & 29 & 16.647 & --- & 0.314 & ZGP-BLAP-06$^{[2]}$ \\
OGLE-BLAP-059 & 18:54:01.82 & $-$06:33:22.2 & 27.38398 & $-$3.55829 & 0.13 & 0.11 & 33.5823049 & 91 & 17.113 & --- & 0.296 & --- \\
OGLE-BLAP-060 & 19:03:29.08 & $-$00:05:02.0 & 34.23148 & $-$2.71840 & 0.16 & 0.09 & 44.6277164 & 118 & 16.836 & --- & 0.109 & ZGP-BLAP-07$^{[2]}$ \\
OGLE-BLAP-061 & 19:13:08.78 & 12:04:51.5 & 46.13835 & 0.75200 & 0.44 & 0.08 & 35.1375078 & 107 & 16.347 & --- & 0.232 & ZGP-BLAP-08$^{[2]}$ \\
\hline
\end{tabular}
}
{\scriptsize \\ References: [1] Pietrukowicz \etal (2017), [2] McWhirter and Lam (2021)}
\end{sidewaystable}

\begin{figure}
\includegraphics[width=12.71cm]{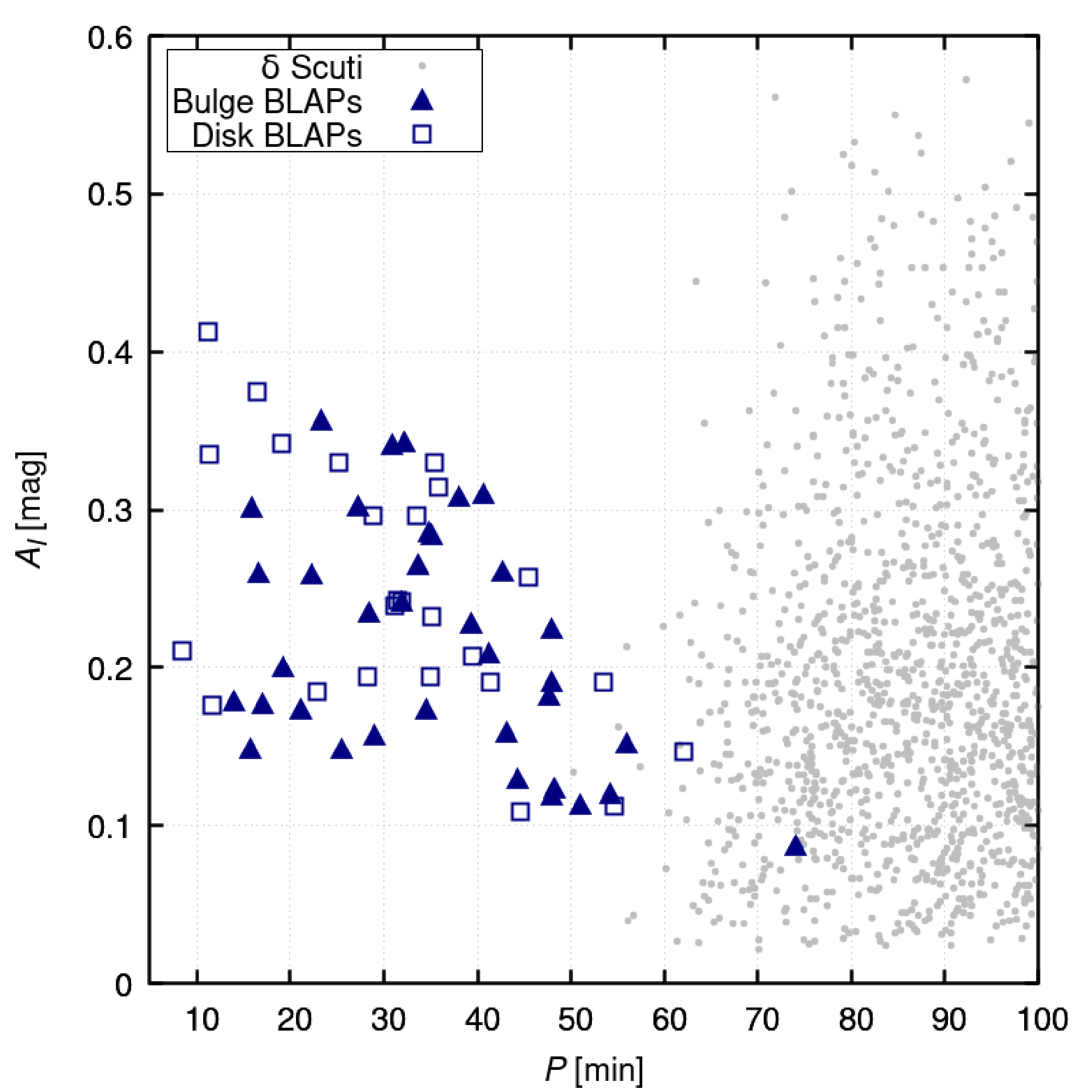}
\FigCap{Period--amplitude diagram for all 61 BLAPs detected in the OGLE fields. Twenty-five stars from the Galactic disk fields (including the prototype object OGLE-BLAP-001) are marked with blue empty squares. The original 13 BLAPs (Pietrukowicz \etal 2017) and 23 additional BLAPs recently found in the Galactic bulge fields (Pietrukowicz \etal 2023, submitted) are marked with blue filled triangles. Positions of $\delta$~Sct stars from the OGLE-IV Galactic disk are represented with gray points.}
\end{figure}

\begin{figure}[h!]
\includegraphics[width=12.71cm]{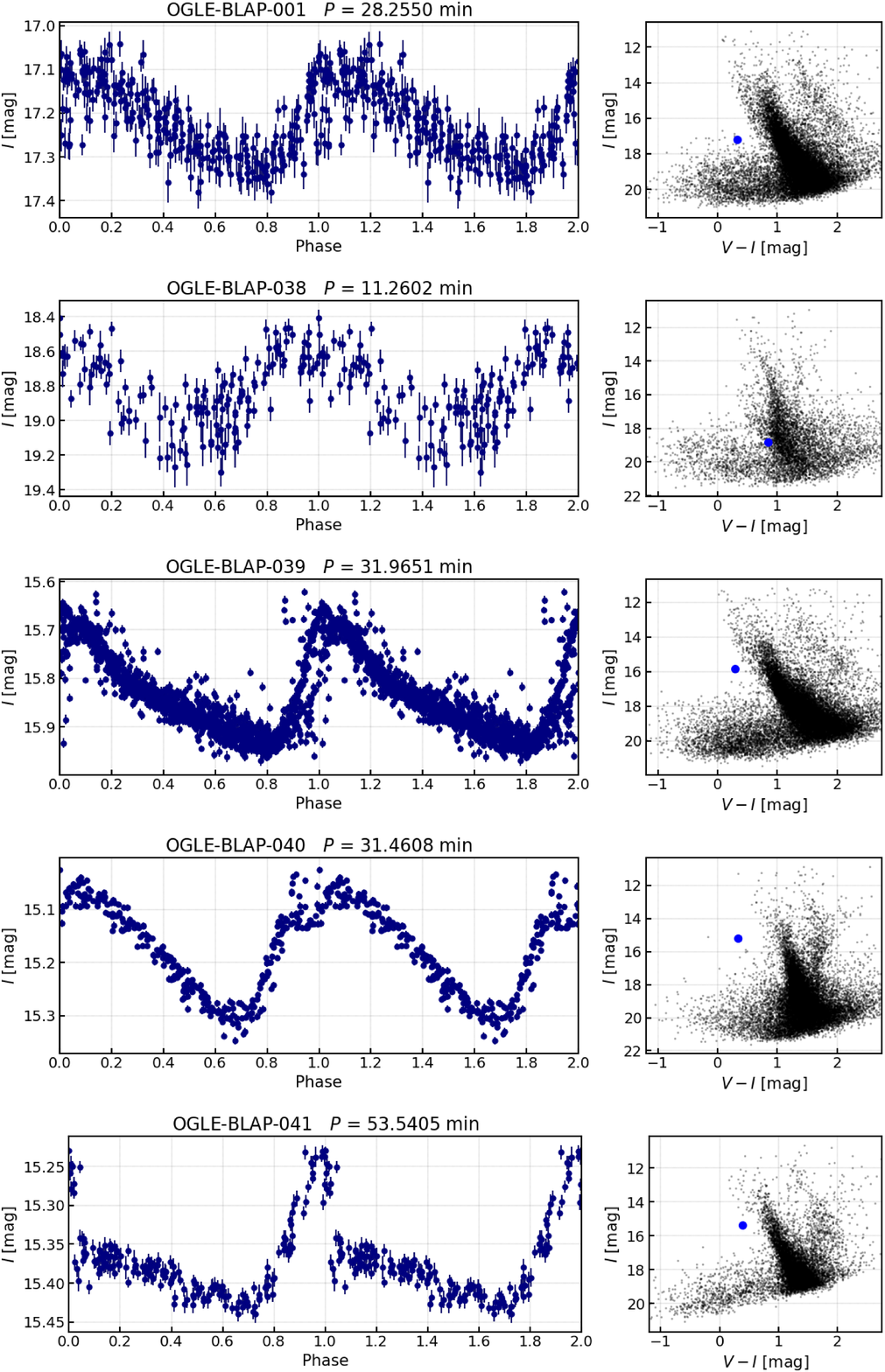}
\FigCap{Phase-folded \textit{I}-band light curves (left panels) and color-magnitude diagrams (right panels) for BLAPs detected in the OGLE-IV Galactic disk fields. The diagrams were constructed for stars from the corresponding OGLE-IV subfield. Positions of the BLAPs are marked with blue dots. A significant period change (as the light curve shift) can be noted in object OGLE-BLAP-039.}
\end{figure}

\begin{figure}[h!]
\includegraphics[width=12.71cm]{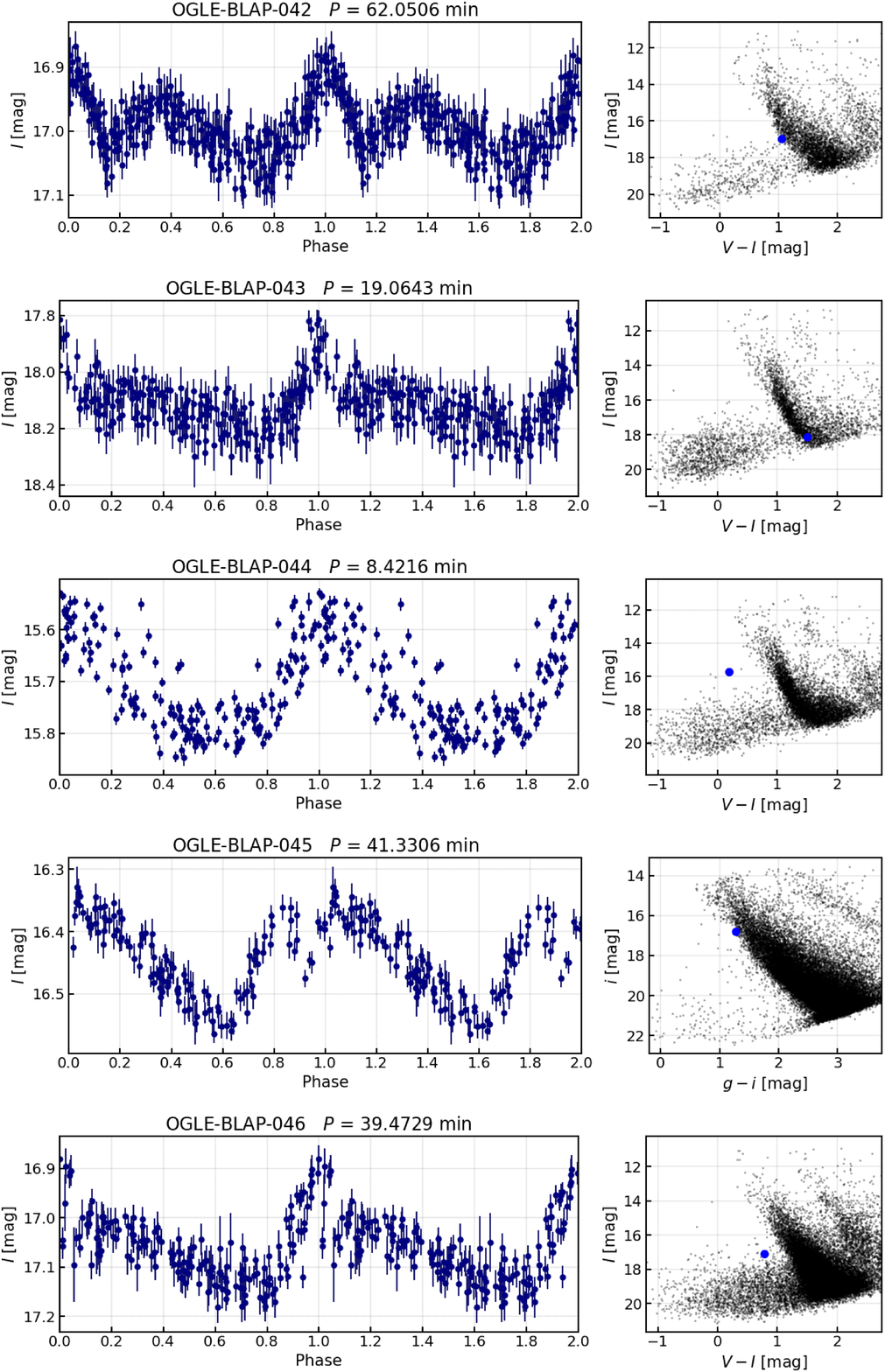}
\FigCap{Same as Fig.~ 3 for another five BLAPs. Light curve blur in object OGLE-BLAP-044 is caused by significant period change. The color-magnitude diagram for OGLE-BLAP-045 is based on data from the DECam Plane Survey.}
\end{figure}

\begin{figure}[h!]
\includegraphics[width=12.71cm]{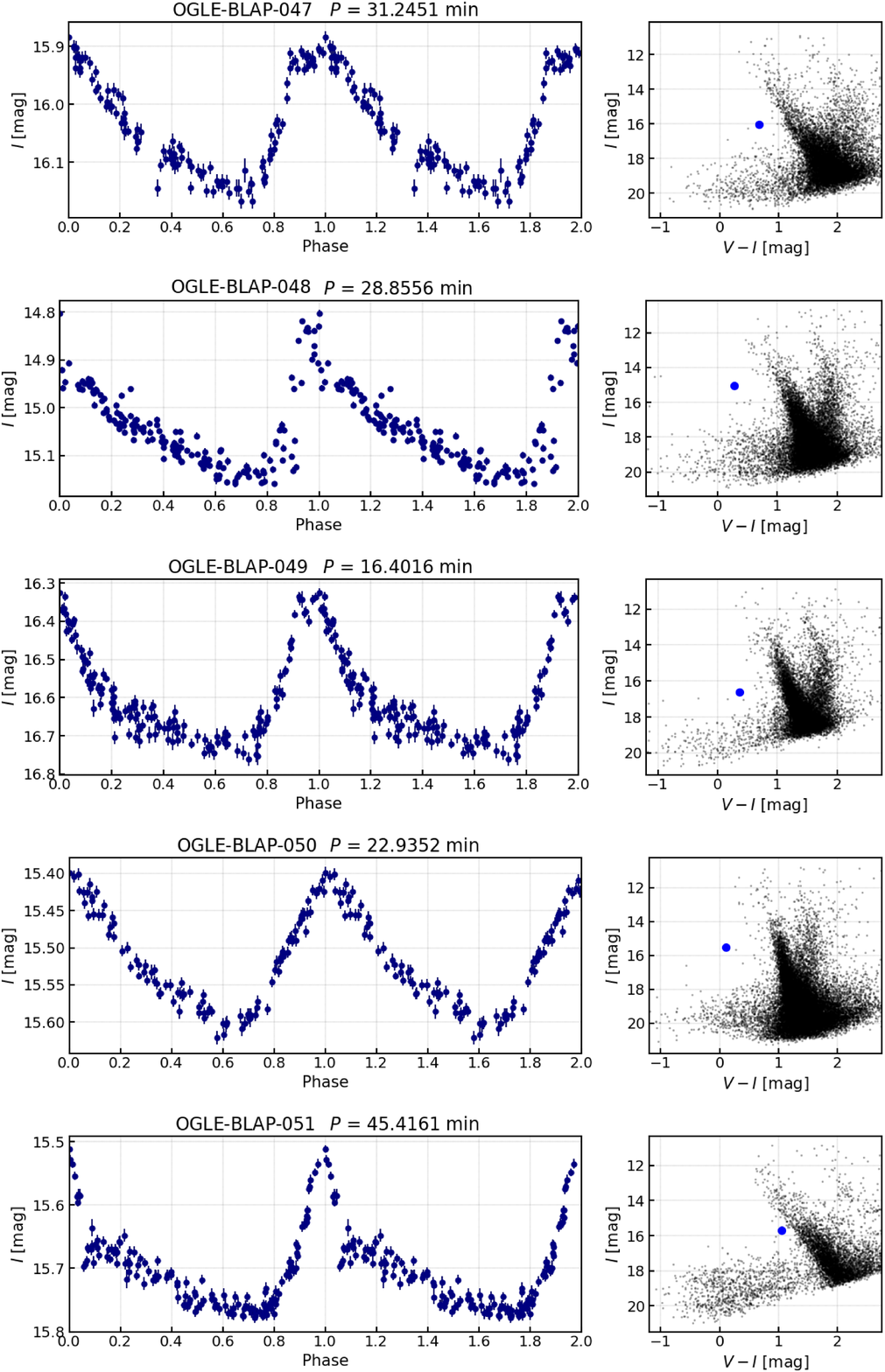}
\FigCap{Same as Figs. 3--4 for another five BLAPs.}
\end{figure}

\begin{figure}[h!]
\includegraphics[width=12.71cm]{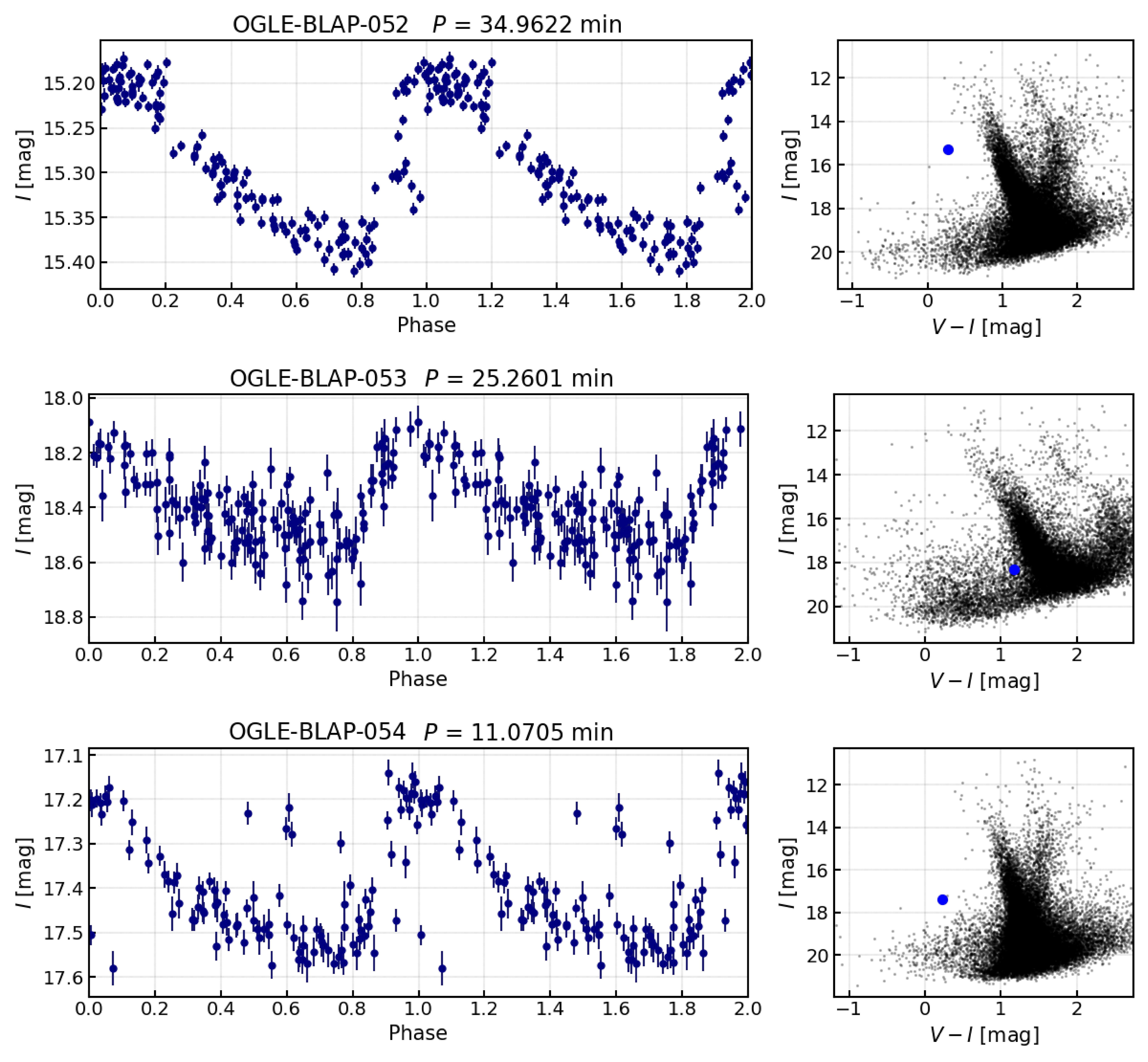}
\FigCap{Same as Figs. 3--5 for the remaining three BLAPs. Points that seem to be outliers in the light curve of OGLE-BLAP-054 result from a large period change.}
\end{figure}

In our sample, star OGLE-BLAP-044 has the shortest pulsation period of 8.42~min, which is the shortest value among all BLAPs detected in the OGLE data so far. The longest period of 62.05~min is observed in OGLE-BLAP-042, which is the second longest known BLAP after OGLE-BLAP-022 with $P=74.05 $ min (Pietrukowicz \etal 2023, submitted). For some of the objects, period changes are clearly noticeable. This can be seen in the case of OGLE-BLAP-039 (for which the majority of observations were taken over three nights), OGLE-BLAP-044, and OGLE-BLAP-055. $I$-band amplitudes in our BLAP sample are in the range of 0.109--0.413~mag. In the light curves of some objects, a bump or dip around the maximum is present, notably in OGLE-BLAP-040, 045, 047, 057, and 060. We noticed that some objects have amplitudes and light curve shapes very similar to each other despite slightly different periods. This refers to OGLE-BLAP-057 ($P = 54.65$~min, $A_{I} = 0.112$~mag) and OGLE-BLAP-060 ($P = 44.63$~min, $A_{I} = 0.109$~mag), if one compares the rich ZTF-$r$ light curves. Object OGLE-BLAP-047 ($P = 31.24$~min, $A_{I} = 0.239$~mag) is an almost perfect analog of star OGLE-BLAP-009 ($P = 31.93$~min, $A_{I} = 0.242$~mag) observed toward the Galactic bulge (Pietrukowicz \etal 2017). There is an unusual scatter around the maximum light in object OGLE-BLAP-040. This object is present in two overlapping OGLE-IV fields. The aforementioned effect is seen in both light curves implying that it is a real intrinsic feature of the star. The new BLAPs show only one periodicity. Based on Gaia EDR3 parallaxes (Gaia collaboration \etal 2021), we can conclude that all the reported here BLAPs are rather distant stars located at several kpc.

\begin{figure}[h!]
\includegraphics[width=12.71cm]{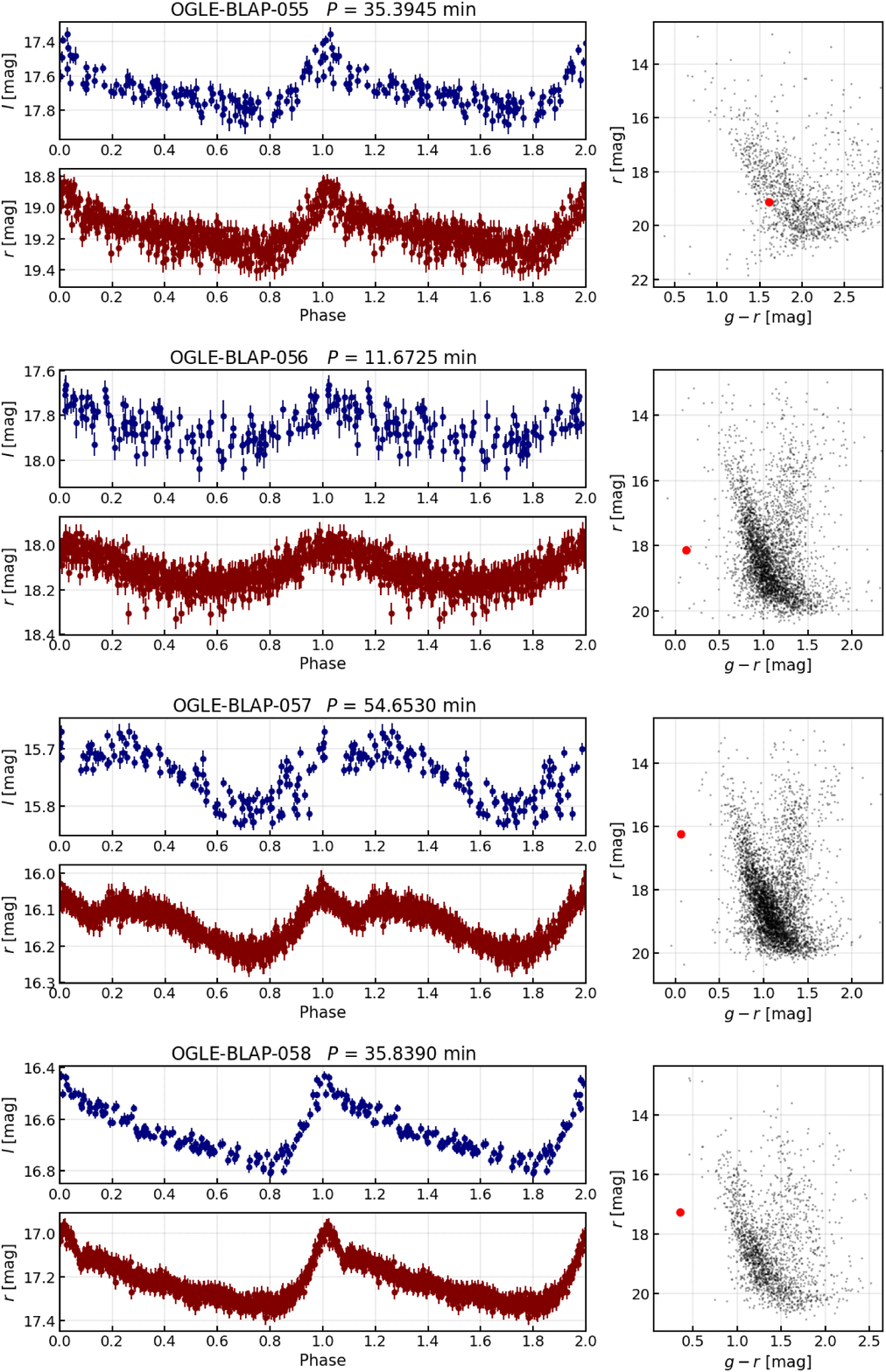}
\FigCap{Phase-folded light curves (left columns) and color-magnitude diagrams (right columns) for BLAPs detected in OGLE-IV Galactic disk fields with positive longitude (DG fields). The light curves plotted in red are based on ZTF \textit{r}-band measurements, whereas blue light curves use OGLE-IV $I$-band data. The diagrams are constructed for stars located within a 10$'\times10'$ box centered on the target BLAPs using only ZTF data.}
\end{figure}

\begin{figure}[h!]
\includegraphics[width=12.71cm]{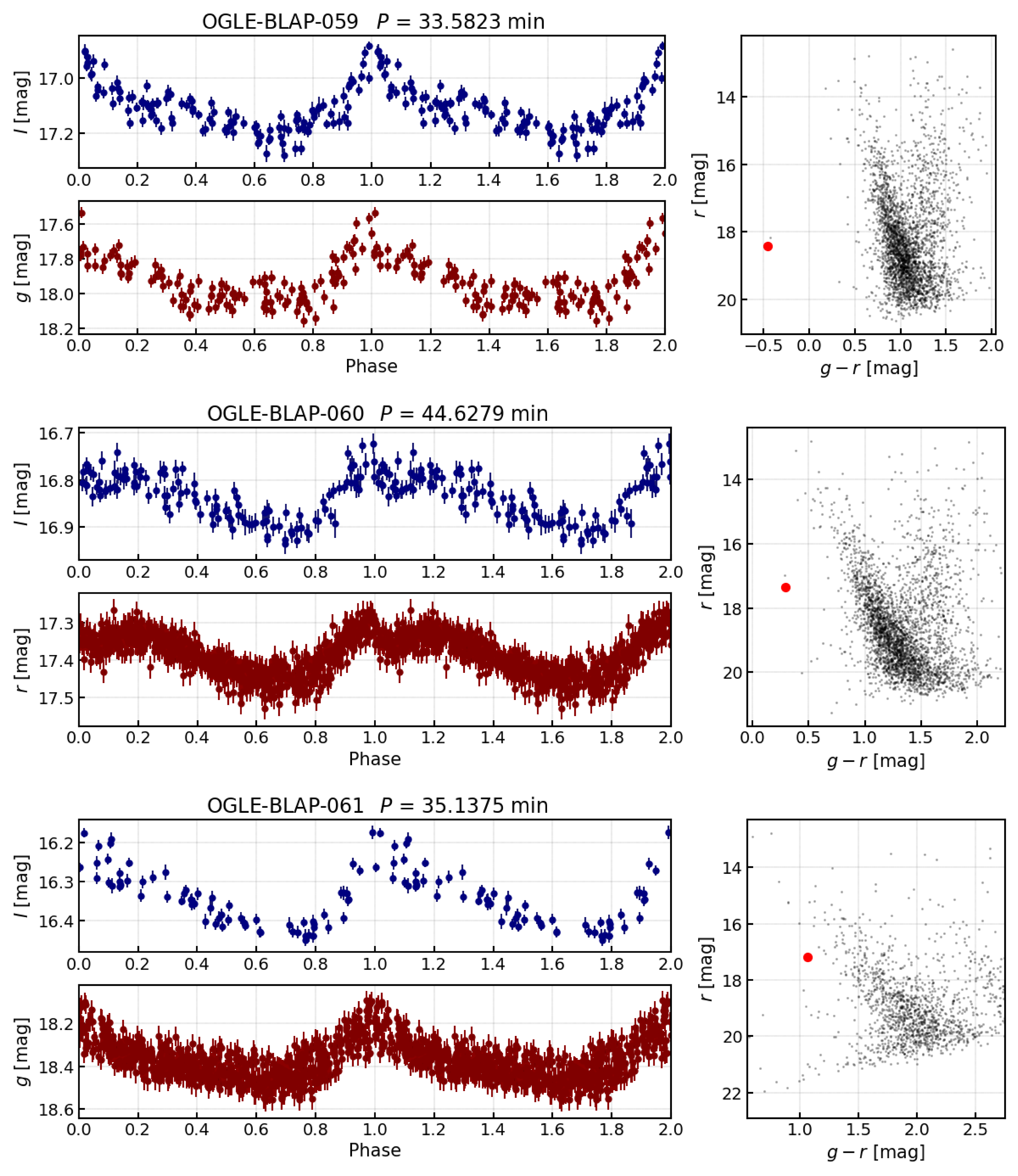}
\FigCap{Same as Fig.~7 for the remaining stars from OGLE-IV DG fields. For the CMDs \textit{g} and \textit{r}-band data from ZTF were used. Light curves of objects OGLE-BLAP-059 and OGLE-BLAP-061 plotted in red are based on ZTF $g$-band data.}
\end{figure}

\subsection{Eclipsing Binary Systems}

Our analysis of the OGLE-IV Galactic disk data led to the identification of six short-period eclipsing binary systems. None of them were known earlier. In Fig.~9, we present the CMD next to the light curve folded with the orbital period for each of the systems, while in Table~2, we provide their photometric properties. All the binaries but object GD1548.32.5823 have the orbital periods $P_{\rm orb}<2$~h. The shortest detected system, GD1468.32.3751, has $P_{\rm orb} = 38.31$ min. This is the only system with the orbital period below 1~h. Objects GD1200.31.11089 and GD1548.32.5823 seem to be contact systems, while the remaining four objects are likely semi-detached systems. Both candidates for contact systems have the orbital periods shorter than the shortest-period object of this type found in the OGLE data so far (OGLE-BLG-ECL-000066, $P = 2.36$~h; Soszy{\'n}ski \etal 2015). Due to the lack of $V$-band measurements, the CMD for GD1200.31.11089 is constructed based on data from the DECam Plane Survey. We found no X-ray counterparts to the discovered systems.

\begin{figure}[h!]
\includegraphics[width=12.71cm]{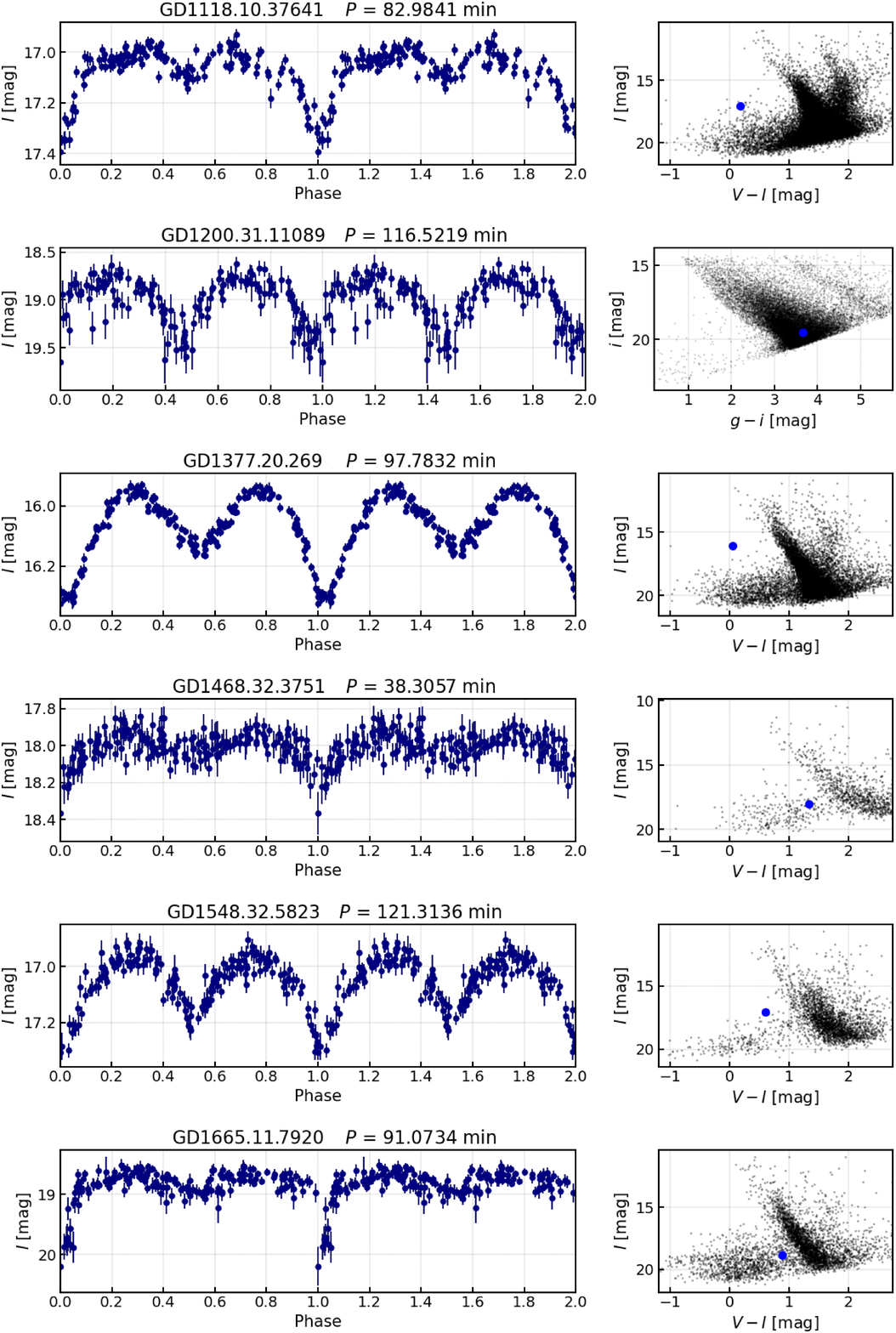}
\FigCap{Phase-folded \textit{I}-band light curves (left panels) and CMDs (right panels) for eclipsing binary systems found in the OGLE-IV Galactic disk fields. The CMD for GD1200.31.11089 is based on data from the DECam Plane Survey.}
\end{figure}

\subsection{Other Short-Period Variable Stars}

In addition to the previous objects, six short-period large-amplitude variable stars were detected (see Table~3). The variables were noticed thanks to peculiar shapes of the light curves, often after doubling the period. All of the stars have period below 2~h. Phase-folded curves and CMDs are presented in Fig.~10. It is challenging to indicate whether their variability is due to binarity or rotational effect, in particular in the case of objects GD1537.05.11659 and GD2077.07.161.  We note that star GD1670.31.3740 has an X-ray counterpart detected by ROSAT satellite (Boller \etal 2016). The source 2RXS J071305.7-084519 is separated by $5\zdot\arcs6$ from the star, which is comparable to the spatial resolution of ROSAT. The detection of X-rays indicates accretion processes and thus binary nature of the source. This object as the only one in the group was found to be variable in Gaia DR3 and classified as a cataclysmic variable (to the "CV" class). Our light curve exhibits a long-term trend, which we subtracted before phasing. The period of 114.61 min, not half of this value, seems to be correct for this type of binary systems. In Fig.~11, we show the time-domain and phase-folded light curve before detrending. Its full amplitude reaches 2.1~mag in $I$. Star GD2117.17.7515 has a very odd light curve shape with an $I$-band amplitude of 0.483~mag. From Gaia EDR3 astrometry, we know that this is a relatively close object ($215\pm9$ pc, Bailer-Jones \etal 2021). The period of 90.75~min and blue color of the object indicate that this is probably another cataclysmic variable. Out of the 12 detected eclipsing binaries and other short-period variables, five objects are located at relatively close distances, below 0.5~kpc. The remaining seven stars are located farther than 1~kpc.

\begin{sidewaystable}
\caption{Photometric properties of detected eclipsing binary systems}
\label{}
{\scriptsize
\begin{tabular}{lcccrcrrccccc}
\hline
\multicolumn{1}{c}{OGLE-IV ID}  & RA (J2000) & Dec (J2000)  & $l$  & \multicolumn{1}{c}{$b$}      & $\varpi$   & $\sigma_\varpi$ & \multicolumn{1}{c}{$\textit{P}_{\rm orb}$}       & $\sigma_P$ & $I$ & $V-I$ & $A_{I}$ & Other name \\
 & [h:m:s] & [${}^\circ:':''$] & [${}^\circ$] & \multicolumn{1}{c}{[${}^\circ$]} & [mas] & [mas] & \multicolumn{1}{c}{[min]} & {\tiny{[$ 10^{-7}$ min]}} & [mag] & \multicolumn{1}{c}{[mag]} &  [mag] & [Ref.] \\ \midrule
GD1665.11.7920  & 07:19:41.47  & $-$19:53:47.3 & 233.79233 & $-$2.99652 & 3.18 & 0.32 & 91.0734004  & 133 & 18.831 & 0.885  & 1.133 & --- \\
GD1548.32.5823  & 08:08:39.16  & $-$32:50:07.9 & 250.46588 & $-$0.00261 & 0.23 & 0.07 & 121.3136403 & 215 & 17.055 & 0.602  & 0.341 & --- \\
GD1468.32.3751  & 09:01:00.75  & $-$48:12:45.4 & 268.68136 & $-$1.27610 & 0.51 & 0.17 & 38.3057050  & 80 & 18.004 & 1.340  & 0.222 & --- \\
GD1377.20.269   & 10:35:38.92 & $-$54:47:29.8 & 284.12262 & 3.04366  & 0.30 & 0.03 & 97.7832499  & 92 & 16.066 & 0.047  & 0.368 & --- \\
GD1200.31.11089 & 15:10:44.42 & $-$58:01:22.2 & 320.53575 & 0.00644 & 3.22 & 1.02 & 116.5218887 & 464 & 19.005 & \multicolumn{1}{c}{---} & 0.543 & --- \\
GD1118.10.37641 & 16:38:39.19 & $-$51:47:40.5 & 334.15607 & $-$3.26420 & 0.50 & 0.09 & 82.9840853  & 207 & 17.073 & 0.179  & 0.338 & --- \\
\hline
\end{tabular}}
%\end{sidewaystable}

%\begin{sidewaystable}
\caption{Photometric properties of other short-period variable stars}
\label{}
{\scriptsize
\begin{tabular}{lcccrcrrccrcc}
\hline
\multicolumn{1}{c}{OGLE-IV ID}  & RA (J2000) & Dec (J2000)  & $l$  & \multicolumn{1}{c}{$b$}      & $\varpi$   & $\sigma_\varpi$ & \multicolumn{1}{c}{$\textit{P}$}       & $\sigma_P$ & $I$ & \multicolumn{1}{c}{$V-I$} & $A_{I}$ & Other name \\
 & [h:m:s] & [${}^\circ:':''$] & [${}^\circ$] & \multicolumn{1}{c}{[${}^\circ$]} & [mas] & [mas] & \multicolumn{1}{c}{[min]} & {\tiny{[$ 10^{-7}$ min]}} & [mag] & \multicolumn{1}{c}{[mag]} &  [mag] & [Ref.] \\ \midrule
GD1670.31.3740  & 07:13:06.14  & $-$08:45:22.1  & 223.18737 & 0.77988  & 2.46 & 0.08 & 114.6144716 & 813 & 16.720 & 0.893  & 2.149 & 2RXS J071305.7-084519$^{[1]}$ \\
GD1537.05.11659 & 08:15:30.78  & $-$34:21:23.4 & 252.52163 & 0.35499  & 0.51 & 0.05 & 73.7073948  & 95 & 16.845 & $-$0.032 & 0.211 & --- \\
GD2117.17.7515  & 08:22:39.46  & $-$42:58:52.7 & 260.43536 & $-$3.35434 & 4.68 & 0.24 & 90.7473704  & 1059 & 18.320 & 1.247  & 0.483 & --- \\
GD2077.07.161   & 09:18:31.83  & $-$60:23:22.6 & 279.39515 & $-$7.65360 & 0.45 & 0.04 & 101.1611974 & 637 & 16.063 & $-$0.008 & 0.206 & --- \\
GD2011.25.1811  & 12:34:18.86 & $-$68:14:56.5 & 301.33966 & $-$5.43022 & 0.57 & 0.16 & 88.0693069  & 1054 & 16.995 & 1.319  & 0.171 & --- \\
GD1072.09.419   & 17:09:09.45 & $-$44:04:03.9 & 343.37511 & $-$2.35751 & 6.76 & 0.13 & 105.9106750 & 660 & 16.995 & 1.207  & 0.201 & --- \\
\hline
\end{tabular}}
{\scriptsize \\ References: [1] Boller \etal (2016)}
\end{sidewaystable}

\begin{figure}
\includegraphics[width=12.71cm]{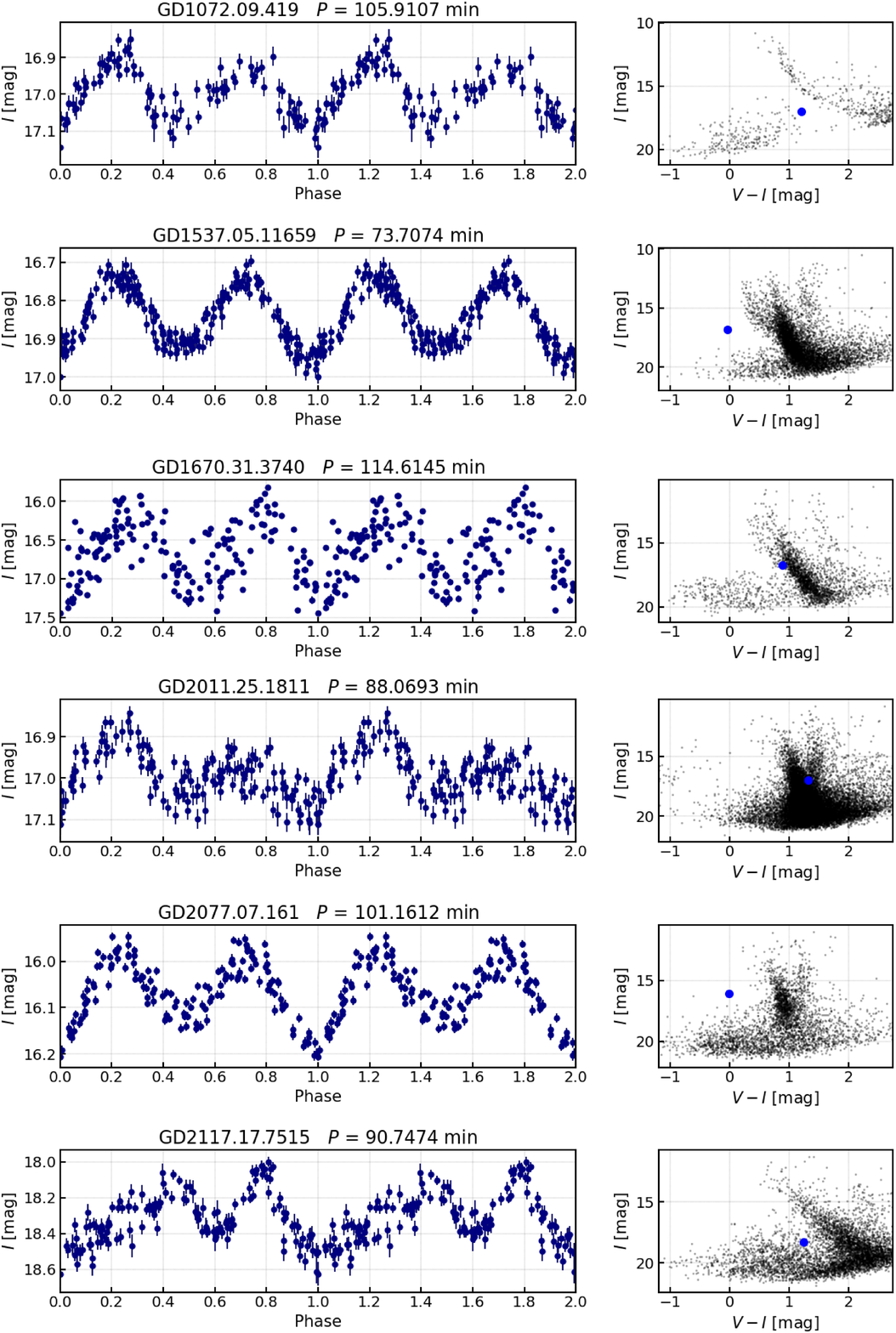}
\FigCap{Other short-period large-amplitude variable objects found in the OGLE-IV Galactic disk fields. The light curve of object GD1670.31.3740 was phase-folded after subtraction of a long-term trend shown in Fig.~11.}
\end{figure}

\begin{figure}
\includegraphics[width=12.71cm]{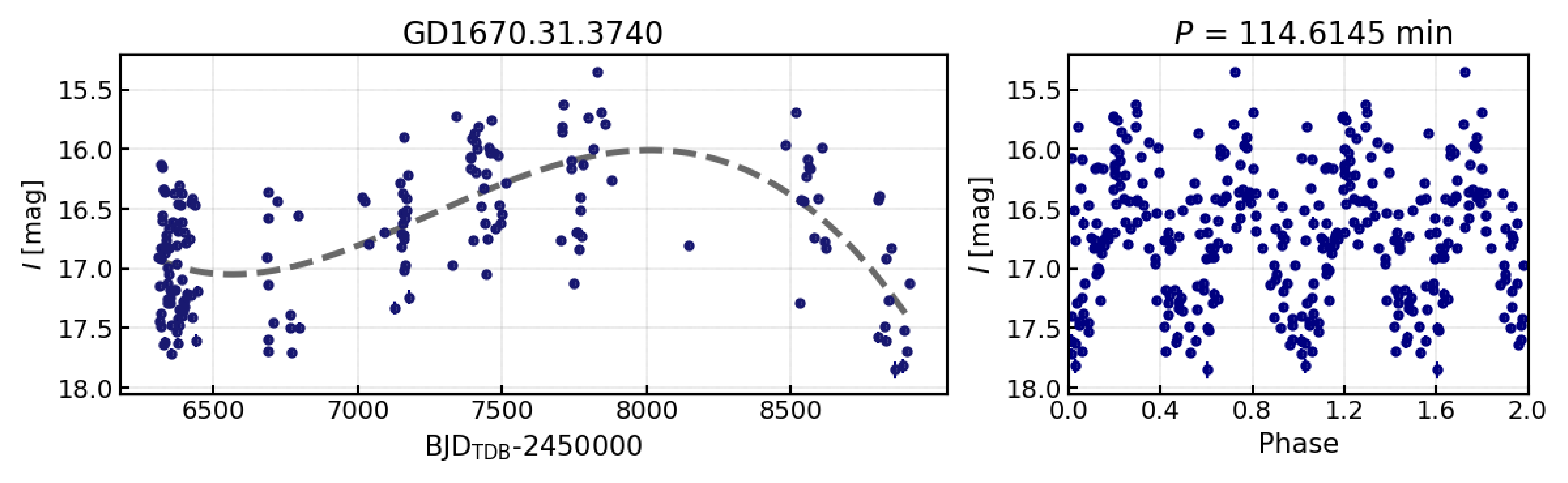}
\FigCap{Time-domain (left panel) and phase-folded (right panel) $I$-band light curve of object GD1670.31.3740 exhibiting a long-term trend. Note a huge amplitude of the brightness changes in this star.}
\end{figure}

\section{Summary}
This study focused on the search for short-period ($<1$~h) variable stars in the OGLE-IV Galactic disk fields covering almost 2400 square degrees, with a specific emphasis on the detection of BLAPs and compact binary systems. The monitoring of about one billion stellar sources allowed the detection of 37 large-amplitude ($>0.1$~mag) variable stars. We identified 25 BLAPs, 20 of which are new findings. Altogether with the very recent detection of 23 objects toward the inner Galactic bulge (Pietrukowicz \etal 2023, submitted), our search increased the number of BLAPs to over 80. We reported the detection of the shortest-period known BLAP in the OGLE area, OGLE-BLAP-044 with $P = 8.42$~min. In addition to the BLAPs, we identified six short-period eclipsing binary systems with orbital periods in the range 38.3--121.3~min. We informed about the detection of additional six short-period large-amplitude variable stars. Two of the stars are likely cataclysmic variables. Follow-up spectroscopic observations would allow their final classification.

\Acknow{We thank all the OGLE observers for their contribution to the collection of the photometric data over the decades. This work has been funded by the National Science Centre, Poland, grant no.~2022/45/B/ST9/00243 to I.S. For the purpose of Open Access, the author has applied a CC-BY public copyright licence to any Author Accepted Manuscript (AAM) version arising from this submission. We used data from the European Space Agency (ESA) mission Gaia, processed by the Gaia Data Processing and Analysis Consortium (DPAC). Funding for the DPAC has been provided by national institutions, in particular the institutions participating in the Gaia Multilateral Agreement.}

\end{document}